\documentclass[twocolumn]{aastex61}

\usepackage{color}
\usepackage{booktabs}

\definecolor{red}{rgb}{1.0,0.0,0.0}
\definecolor{blue}{rgb}{0.0,0.5,1.0}

\received{November 7, 2018}
\revised{January 11, 2019}
\accepted{January 11, 2019}
\submitjournal{ApJ}

\shorttitle{Imaging of the AR\,Pup disk}
\shortauthors{Ertel et al.}
\watermark{DRAFT}

\begin{document}

\title{Resolved imaging of the AR\,Puppis circumbinary disk\footnote{Based on observations made with ESO Telescopes at the La
Silla Paranal Observatory under program IDs~097.D-0385 (SPHERE) and 094.D-0865 (PIONIER).}}

\correspondingauthor{Steve Ertel}
\email{sertel@email.arizona.edu}

\author{S.~Ertel}
\affiliation{Large Binocular Telescope Observatory, 933 North Cherry Avenue, Tucson, AZ 85721, USA}
\affiliation{Steward Observatory, Department of Astronomy, University of Arizona, 993 N. Cherry Ave, Tucson AZ, 85721, USA}
\affiliation{European Southern Observatory, Alonso de C\'ordova 3107, Vitacura, Casilla 19001, Santiago 19, Chile}

\author{D.~Kamath}
\affiliation{Research School of Astronomy and Astrophysics, Australian National University, Cotter Road, Weston Creek ACT 2611,
Australia}
\affiliation{Department of Physics and Astronomy, Macquarie University, Sydney, NSW 2109, Australia}
\affiliation{Instituut voor Sterrenkunde, K.U.Leuven, Celestijnenlaan 200D bus 2401, B-3001 Leuven, Belgium}

\author{M.~Hillen}
\affiliation{Instituut voor Sterrenkunde, K.U.Leuven, Celestijnenlaan 200D bus 2401, B-3001 Leuven, Belgium}

\author{H.~van~Winckel}
\affiliation{Instituut voor Sterrenkunde, K.U.Leuven, Celestijnenlaan 200D bus 2401, B-3001 Leuven, Belgium}

\author{J.~Okumura}
\affiliation{Centre for Astrophysics, University of Southern Queensland, West Street, Toowoomba, QLD 4350, Australia}

\author{R.~Manick}
\affiliation{Instituut voor Sterrenkunde, K.U.Leuven, Celestijnenlaan 200D bus 2401, B-3001 Leuven, Belgium}

\author{H.~M.~J.~Boffin}
\affiliation{European Southern Observatory, Karl-Schwarzschild-Strasse 2, D-85748 Garching bei M\"unchen, Germany}

\author{J.~Milli}
\affiliation{European Southern Observatory, Alonso de C\'ordova 3107, Vitacura, Casilla 19001, Santiago 19, Chile}

\author{G.~H.-M.~Bertrang}
\affiliation{Max Planck Institute for Astronomy, K\"onigstuhl 17, D-69117 Heidelberg, Germany}

\author{L.~Guzman-Ramirez}
\affiliation{Leiden Observatory, Leiden University, Niels Bohrweg 2, 2333 CA Leiden, The Netherlands}
\affiliation{European Southern Observatory, Alonso de C\'ordova 3107, Vitacura, Casilla 19001, Santiago 19, Chile}

\author{J.~Horner}
\affiliation{Centre for Astrophysics, University of Southern Queensland, West Street, Toowoomba, QLD 4350, Australia}

\author{J.~P.~Marshall}
\affiliation{Academia Sinica Institute of Astronomy and Astrophysics, 11F of AS/NTU Astronomy-Mathematics Building, No.1,
Section 4, Roosevelt Road, Taipei 10617, Taiwan, ROC}

\author{P.~Scicluna}
\affiliation{Academia Sinica Institute of Astronomy and Astrophysics, 11F of AS/NTU Astronomy-Mathematics Building, No.1,
Section 4, Roosevelt Road, Taipei 10617, Taiwan, ROC}

\author{A.~Vaz}
\affiliation{Steward Observatory, Department of Astronomy, University of Arizona, 993 N. Cherry Ave, Tucson, AZ, 85721, USA}

\author{E.~Villaver}
\affiliation{Universidad Aut\'onoma de Madrid, Departamento de F\'isica Te\'orica, E-28049 Madrid, Spain}

\author{R.~Wesson}
\affiliation{Department of Physics and Astronomy, University College London, Gower Street, London WC1E 6BT, UK}
\affiliation{European Southern Observatory, Alonso de C\'ordova 3107, Vitacura, Casilla 19001, Santiago 19, Chile}

\author{S.~Xu}
\affiliation{Gemini Observatory, 670 N. A'ohoku Place, Hilo, HI 96720, USA}
\affiliation{European Southern Observatory, Karl-Schwarzschild-Strasse 2, D-85748 Garching bei M\"unchen, Germany}



\begin{abstract}
Circumbinary disks are common around post-asymptotic giant branch (post-AGB) stars with a stellar companion on orbital time scales
of a few 100 to few 1000 days. The presence of a disk is usually inferred from the system's spectral energy distribution (SED),
and confirmed, for a sub-sample, by interferometric observations. We used the Spectro-Polarimetric High-contrast Exoplanet
REsearch (SPHERE) instrument on the Very Large Telescope to obtain extreme adaptive optics assisted scattered light images
of the post-AGB binary system AR\,Puppis. Data have been obtained in the $V$, $I$, and $H$~bands. Our observations have produced
the first resolved images of AR\,Puppis' circumbinary disk and confirm its edge-on orientation. In our high angular-resolution
and high dynamic-range images we identify several structural components such as a dark mid-plane, the disk
surface, and arc-like features. We discuss the nature of these components and use complementary photometric monitoring to relate
them to the orbital phase of the binary system. Because the star is completely obscured by the disk at visible wavelengths, we
conclude that the long-term photometric variability of the system must be caused by variable scattering, not extinction, of star
light by the disk over the binary orbit. Finally, we discuss how the short disk life times and fast evolution of the host stars
compared to the ages at which protoplanetary disks are typically observed make systems like AR\,Puppis valuable extreme laboratories
to study circumstellar disk evolution and constrain the time scale of dust grain growth during the planet formation process.

\end{abstract}




\section{Introduction} \label{sect_intro}

Low- and intermediate-mass stars (around 1 to 8\,$M_\odot$) have their nuclear-burning lives terminated by stellar wind mass
loss on the Asymptotic Giant Branch (AGB) if they are single stars, or possibly by mass transfer to a companion if they are
binary stars.  When the hydrogen-rich envelope of the AGB star becomes very small due to mass loss, the star can no longer
maintain its large radius and it shrinks.  The shrinking causes an increase in $T_{\text{eff}}$ as the object evolves through
the post-Asymptotic Giant Branch (post-AGB) phase, up to $T_{\text{eff}} \sim 3\times10^4$\,K  (see \citealt{vanwi03} for a
review), followed by the planetary nebula and white dwarf phases.  Binarity is thought to be the main driver for breaking the
initially largely spherical envelopes of AGB stars to form bipolar protoplanetary and planetary nebulae \citep{sok04,
sok06, nor06, demar09, bof12}.  Some
binary stars will already interact when the primary expands during its evolution on the Red Giant Branch (RGB).  Such systems,
the dusty post-RGB stars, have recently been identified in the Magellanic Clouds \citep{kam16, man18}.  However, due to
observational limitations, the binary nature and orbital parameters of these systems are yet to be fully explored.

Many close post-AGB binaries display disk-type spectral energy distributions (SEDs).  These are characterized by a clear near
infrared excess indicating that a fraction of the expelled circumstellar dust and gas must be located close to the central
star, near the sublimation radius \citep{vanwi03}.  It is now well established that this feature in the SED indicates the
presence of a stable, compact, Keplerian circumbinary disk, and these sources are referred to as disk sources (e.g.,
\citealt{deruy06, hil14, kam14, kam15}). All non-pulsating Galactic post-AGB disk sources are proven to be binaries with
binary orbital
time scales of the order of one to several years and therefore this specific characteristic of the SED is closely related to the
binary nature of the central star \citep{vanwi09, gez15}. Many pulsating objects were also proven to be binaries (e.g.,
\citealt{gor14, man17}), however, this is more difficult to prove as large amplitude pulsations (e.g., of RV\,Tauri stars) induce
large radial velocity variations. Another interesting property common to most post-AGB objects with disk-type SEDs is a peculiar
photospheric composition depleted of refractory elements (\citealt{gez15, oom18}, Kamath \& Van Winckel 2018, subm., and references
therein).

Initially, the study of the morphology and evolution of post-AGB binaries and their disks was limited to
the analysis of spatially unresolved data of the star and infrared excess from the circumstellar dust. From this, disk sizes
(outer radii) of a few 100\,AU to 1,000\,AU were inferred \citep{deruy06}.  CO rotational mapping has confirmed the Keplerian
rotation in a sample of systems \citep{buj13a} and succeeded to resolve the large scale gas disks and their kinematics in several
systems \citep{buj03, buj13b, buj15, buj16, buj17, buj18}.  Only one similar disk around the central star of the Red Rectangle
(HD\,44179) could be imaged in scattered light with the Hubble Space Telescope \citep{ost97, coh04}.  The first spatially resolved
size measurements of the inner dust disks in other systems were
obtained using optical long baseline interferometry at infrared wavelengths \citep{der06}. The same technique provided a more
detailed view of selected disks \citep{der07, hil14, hil15} and basic size measurements of the disks in 19 systems \citep{hil17}.
The inner rim of the disk in the IRAS\,08544-4431 post-AGB system, as well as its inner binary, were resolved in the $H$-band using
interferometric imaging with VLTI/PIONIER \citep{hil16}. However, interferometric data are unable to provide the high dynamic
range and image fidelity that direct imaging can provide. 

Disks around two less evolved giant stars have recently been imaged using VLT/SPHERE:  L$_2$\,Pup is an AGB star
\citep{ker15} and BP\,Psc is a first ascent giant star \citep{zuc08, deboe17, gaia18}.  We discuss these cases and compare them to the
post-AGB binary disks in Sect.~\ref{sect_comp_disks}.

In this paper, we present the first spatially resolved images of the disk around the post-AGB star AR\,Pup. The SED of this system
suggested a disk orientation close to edge-on \citep{hil17} which is a favorable case for direct imaging because the disk blocks
the direct star light.  We observed the star
for our pilot program to demonstrate the feasibility of resolved imaging of post-AGB binary disks in the visible using extreme
adaptive optics (AO) systems on 8-m-class telescopes. In Sect.~\ref{sect_target} we present a brief summary of the current
literature on AR\,Pup, relevant to this study. We discuss our observing strategy and data reduction in Sect.~\ref{sect_sphere}.
The resulting images are presented and the observed structural components of the disk are discussed in Sect.~\ref{sect_images}.
In Sect.~\ref{sect_discuss} we discuss the connection of our imaging results with other observables of the system, specifically
the shape of its SED and the photometric variability.  In Sect.~\ref{sect_disk_evol} we highlight the prospects of using post-AGB
binary disks as a laboratory to studying circumstellar disk evolution  --  in particular dust grain growth  --  and discuss
them in the context of potential scenarios of second generation planet formation on the stellar post-main sequence. We present
our conclusions in Sect.~\ref{sect_conc}.

\section{Target details}
\label{sect_target}

AR\,Pup is a well studied post-AGB binary system \citep{pol96, kis17, hil17}. The study by \citet{pol96} showed that it is a
RV\,Tauri\footnote{RV\,Tauri stars are luminous, variable stars (Type II Cepheids) with alternating deep and shallow
brightness minima. The RVb subclass exhibits an additional long-term variation.} variable of the photometric subclass RVb.
It has a pulsation period of 76.66\,days and a RVb type modulation period of 1194\,days \citep{kis17}. The latter phenomenon
is commonly attributed to variable scattering and/or line of sight extinction of the star light due to the binary's orbital
motion \citep{wae91, wae96} and thus constrains its orbital period (Manick et al., in prep.).
The system's distance is $\sim$1~to 6\,kpc \citep{vanle07, gaia18} and thus representative of nearby post-AGB stars
\citep{deruy06}. AR\,Pup has been classifed as a G-type star based on the spectroscopically derived stellar parameters from
\citet{gon97},  who find \mbox{$T_{\text{eff}} = 6000$\,K}, \mbox{$\log g$ = 1.5} (cgs system), \mbox{and a [Fe/H] = -1.0}. A
detailed chemical abundance analysis has shown that AR\,Pup displays a photospheric chemistry depleted of refractory elements
and is considered to be a strongly depleted post-AGB star \citep{gez15}.

\begin{figure}
 \centering
 \includegraphics[angle=0,width=\linewidth]{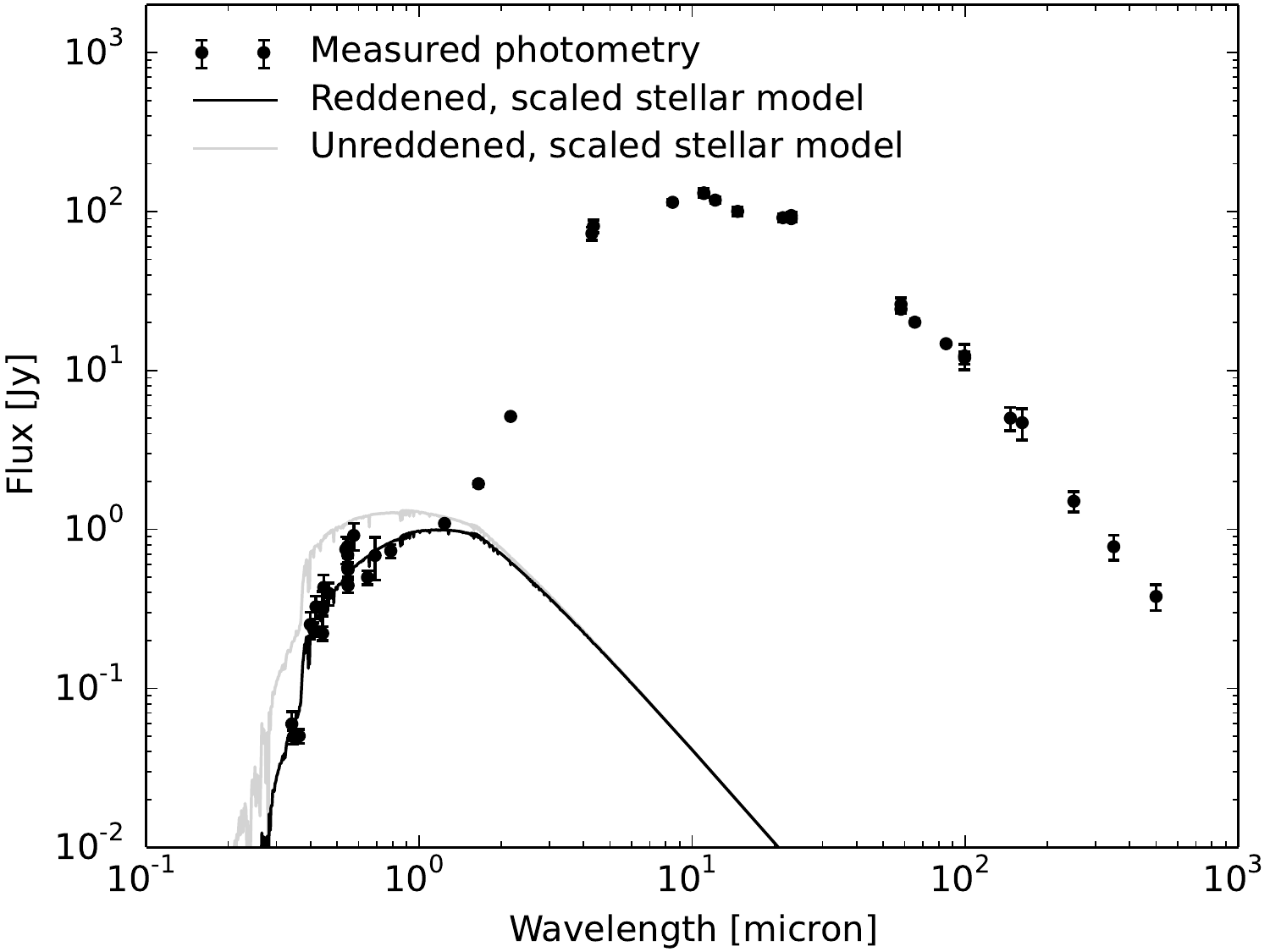}
 \caption{Photometry of AR\,Pup from the literature. A reddened and scaled stellar model photosphere and its de-reddened
  version are over-plotted.  Note that this model does not illustrate the total stellar emission but rather represents the
  amount of scattered star light observed at visible wavelengths (see Sect.~\ref{sect_observables} for a detailed discussion).}
 \label{fig_sed}
\end{figure}

In Fig.~\ref{fig_sed} we show selected photometry of the AR\,Pup system (see Appendix~\ref{app_sed} for a table), adding
measurements from the
Herschel SPIRE Point Source Catalog \citep{schu17} to the data analyzed in detail by \citet{hil17}.  To
guide the eye, we overplot an ATLAS9 model atmosphere \citep{cas04} for the stellar parameters of AR\,Pup, reddened with
$E(B-V) = 0.27$ and scaled to match the visible photometry.  A visual inspection clearly reveals
the presence of a near-infared excess, as typically seen for disk-type sources
(Sect.~\ref{sect_intro}). The disk-type SED-based classification is further confirmed by the position of AR\,Pup on the WISE
color-color plot \citep{gez15}.  The system's variability is obvious from the scatter of the various measurements at similar
wavelengths in the visible.  \citet{hil17} found that the total infrared luminosity is higher than that derived from the
dereddened optical fluxes.  They concluded that the disk of AR\,Pup must be oriented close to edge-on, so that the visible
component of the SED is likely dominated by scattering, with most direct star light being blocked by the optically thick disk.

\section{SPHERE data}
\label{sect_sphere}

The focus of this work is on high angular-resolution observations of AR\,Pup at visible and near-infrared wavelengths.
These data were obtained with the extreme adaptive optics (AO) instrument SPHERE (Spectro-Polarimetric High-contrast
Exoplanet REsearch, \citealt{beu08}) using its cameras ZIMPOL (Zurich Imaging Polarimeter, \citealt{tha08}) in the $V$
and $I$ bands, IRDIS (Infra-Red Dual-beam Imaging and Spectroscopy, \citealt{doh08}) in the $H$ band, and the IFS
(integral field spectrograph, \citealt{cla08}) in the $Y$ to $J$ band. In this section we describe the observations and
data reduction.

\begin{figure*}
 \centering
 \includegraphics[angle=0,width=\linewidth]{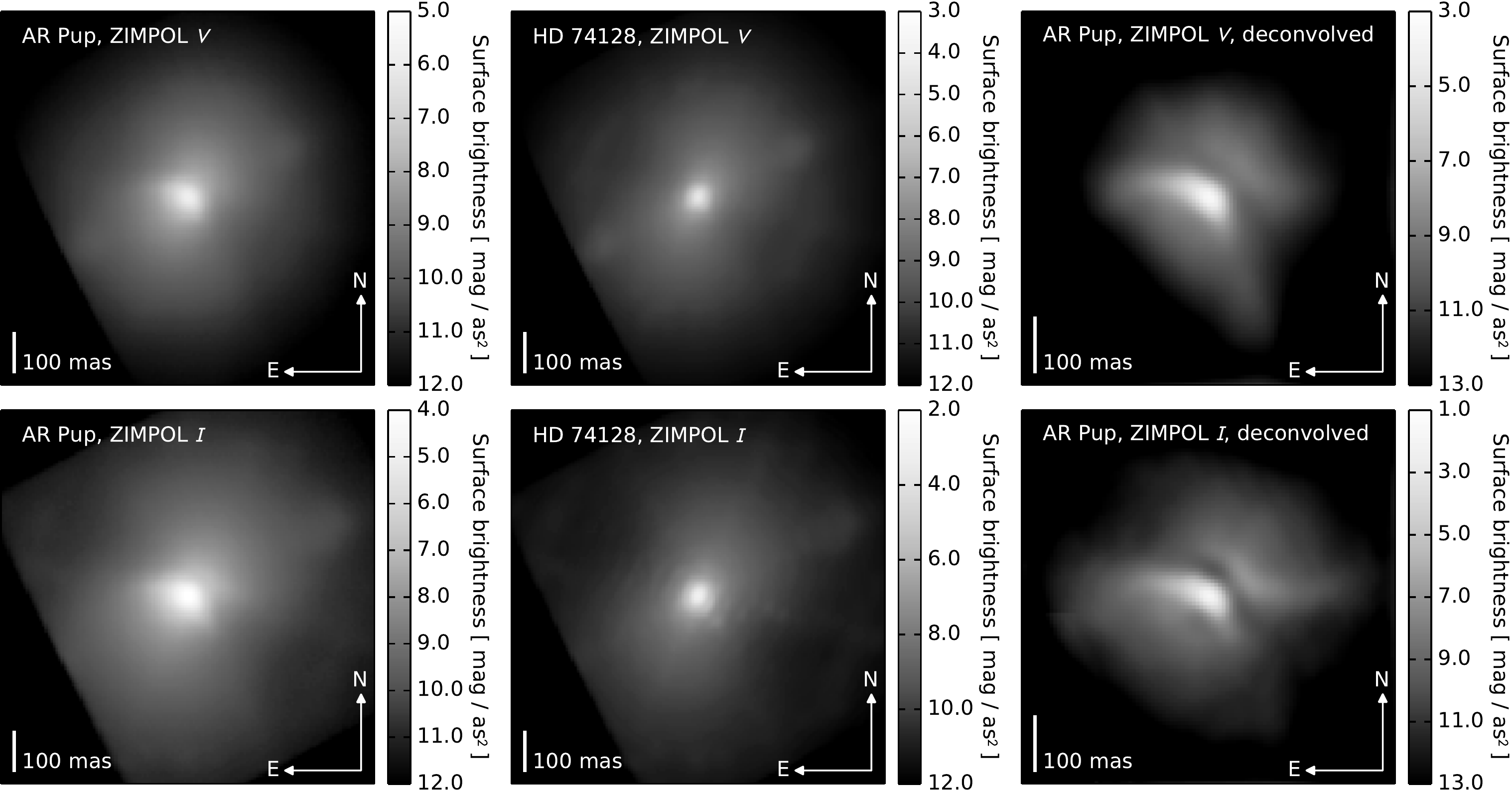}
 \caption{ZIMPOL observations of AR\,Pup (\emph{left}), the PSF reference star HD\,74128 (\emph{center}), and the
  deconvolved images of AR\,Pup (\emph{right}) in $V$ (\emph{top}) and $I$ bands (\emph{bottom}).}
 \label{fig_zimpol_obs}
\end{figure*}

\subsection{Observations}
Observations were carried out on 2016~Apr~4 in service mode (project ID: 097.D-0385, PI: S. Ertel). Observing
conditions were well suited for our high angular-resolution imaging of the relatively bright target ($R\sim9.5$) with a
seeing between $0\farcs8$ and $1\farcs1$ and thin cirrus. The ZIMPOL observations were executed first, followed by the
IRDIS and IFS observations  in IRDIFS mode which combines the two modes
for simultaneous use. In both cases the observations were
paired with observations of a point spread function (PSF) reference star (HD\,74128 for ZIMPOL, observed after AR\,Pup,
$R=9.2$, HD\,75363 for IRDIS, observed before AR\,Pup, $R=7.9$). Both stars were chosen to have a similar brightness as
AR\,Pup in the observing bands used so that the same observing setup (in particular exposure time) as for the science
target could be used. They also have similar brightness in the $R$ band and are close to AR\,Pup in the sky, so that the
AO performance on science target and reference was as similar as possible. All observations were carried out in pupil
stabilized mode to ensure that aberrations were as stable as possible.

For the \textbf{ZIMPOL} observations of both science target and PSF reference star, a single nine point dither pattern
was executed. The number of exposures per dither point was 16 for the science target, and 7 for the reference star.
A detector integration time (DIT) of 4\,s was used for science target and reference star making sure the source counts
were within the linearity regime. This resulted in a total of 144 exposures with a total integration time
of 576\,s on the science target and 63 exposures with a total integration time of 252\,s on the PSF reference star. The
observations were performed simultaneously in the V ($\lambda_0 = 554.0\,\text{nm}$, $\Delta\lambda = 80.6\,\text{nm}$)
and N\_I ($\lambda_0 = 816.8\,\text{nm}$, $\Delta\lambda = 80.5\,\text{nm}$) filters using the two detectors of ZIMPOL
in classical imaging mode (no polarization information obtained). A field stop was unintentionally inserted during the
observations limiting the field of view (FoV) of our ZIMPOL observations to $1''$.

The \textbf{IRDIS} data were obtained in dual band imaging mode in the H2 ($\lambda_c = 1588.8\,\text{nm}$,
$\Delta\lambda = 53.1\,\text{nm}$) and H3 ($\lambda_c = 1667.1\,\text{nm}$, $\Delta\lambda = 55.6\,\text{nm}$) filters 
simultaneously. The observing strategy was analogous to that for ZIMPOL. A 16 point dither pattern was used with three
and two exposures per point on the science target and PSF reference star, respectively, and a DIT of 16\,s. This
results in a total of 48 exposures with a total exposure time of 786\,s on the science target and 32 exposures with a
total exposure time of 512\,s on the PSF reference star.

\textbf{IFS} observations in the $Y$ to $J$ band were carried out in parallel with the IRDIS observations using the 
IRDIFS mode. A total integration time of 1024\,s (DIT=32\,s, NDIT=16, NEXP=2) on the science target and 544\,s
(DIT=32\,s, NDIT=17, NEXP=1) on the reference star were used.

\subsection{Data reduction}
Data reduction was performed using the SPHERE pipeline version 0.18.0 and our own \textit{Python} scripts.

All \textbf{ZIMPOL} frames of both science and reference observations were preprocessed using the pipeline to extract
the informative component of the two detector frames (cutting off prescan and overscan areas and removing every second
row which is masked on the detector) and to determine the overscan bias levels. In the resulting frames, the overscan
bias was subtracted and each odd pixel was averaged with the following pixel in a row, creating an effective pixel
scale of 7.2\,mas. Flat fielding and bad pixel correction were not deemed necessary given the image quality and the
dither pattern performed during the observations. The reference frames were then centered on the source position (the
location of the peak brightness in our images of the target) with
an accuracy of 0.1\,pix and stacked. The science frames were centered, derotated, and stacked. For each science frame,
the stacked reference image was duplicated and rotated by the same angle, so that stacking these rotated PSFs results
in the same rotational smearing of the resulting reference PSF as the derotation of the science frames.

Finally, the resulting science image was deconvolved with the reference PSF using the Richardson-Lucy deconvolution
\citep{ric72, luc74} which converged within 100 iterations. The results of our ZIMPOL data reduction are shown in
Fig.~\ref{fig_zimpol_obs}.

\begin{figure*}
 \centering
 \includegraphics[angle=0,width=\linewidth]{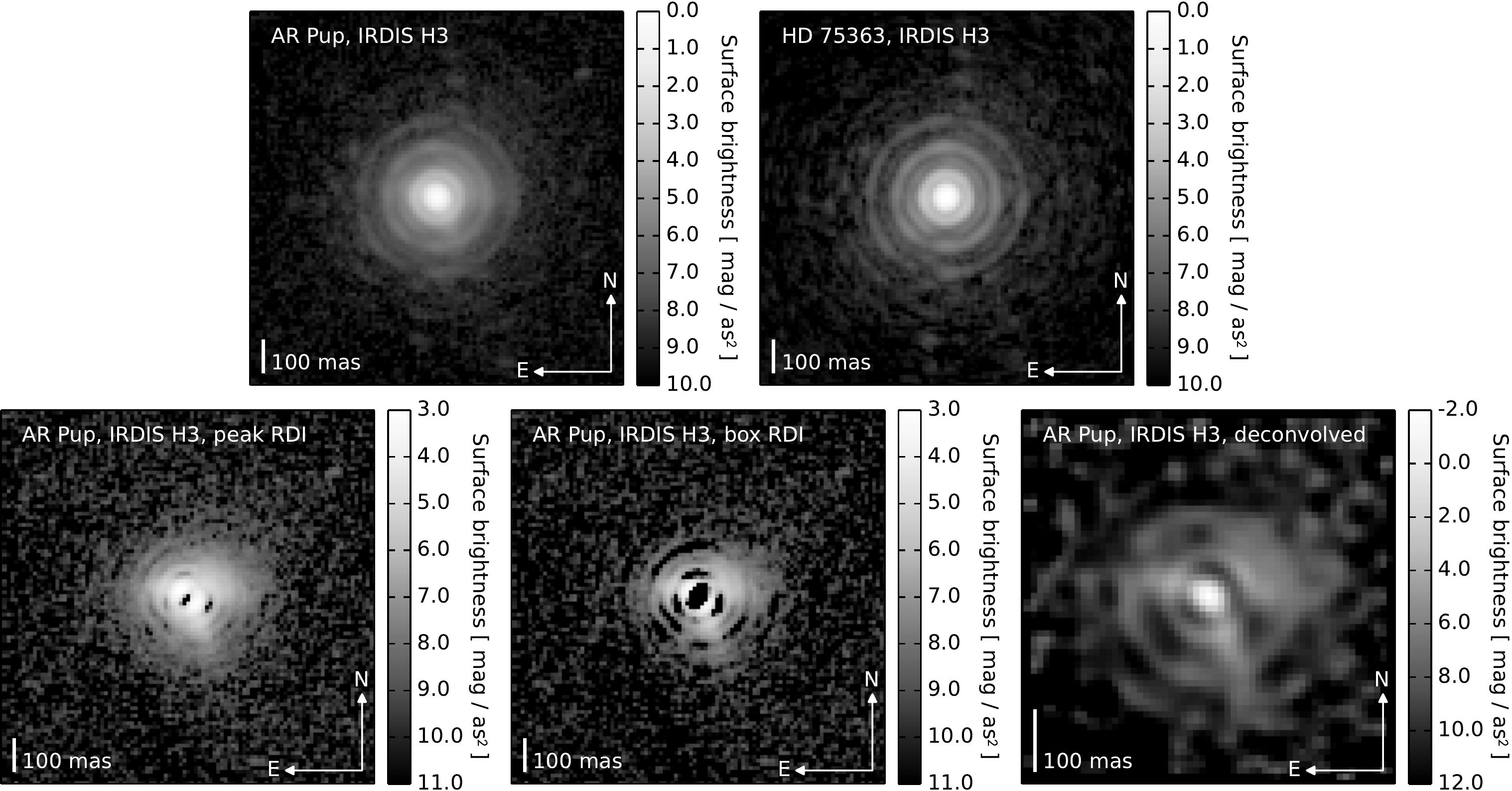}
 \caption{IRDIS observations of AR\,Pup (\emph{top left}), the PSF reference star HD\,75363 (\emph{top right}), and the
  RDI subtracted and deconvolved images of AR\,Pup in the H3 filter (\emph{bottom row}). Both approaches for the RDI
  are shown, where for `peak RDI' the reference star image was scaled to the same peak counts as the AR\,Pup image and
  for `box RDI' it was scaled to have the same counts as the AR\,Pup image in a quadratic aperture centered on the
  peak.}
 \label{fig_irdis_obs}
\end{figure*}

Master background and flat field frames for the \textbf{IRDIS} data were created using the pipeline and the observatory
provided calibration frames. These frames were used to perform standard dark subtraction and flatfielding. In the
resulting frames, bad and hot pixels were corrected using a median filtered map of each frame and a sigma clipping
approach. Finally, to further reduce large scale detector cosmetics (stripes), we subtracted from each column and row
its respective median in regions where no significant source flux was present. The frames still contained two images of
the source -- one for each filter. They were then cut to produce one frame per filter and these frames were centered on
the source. From here the reduction was performed analogous to the ZIMPOL reduction. Reference frames were stacked,
science frames were derotated and stacked, the reference frames were duplicated, rotated the same way as the science
frames, and the results combined to produce a rotationally smeared reference PSF. The counts in the H3 filter are by
factors of $\sim$2 and $\sim$2.5 larger for the calibrator and science target, respectively, compared to the H2 filter.
Thus, combining the two images did not improve the signal-to-noise ratio (S/N) compared to using the H3 filter
alone and we used only the H3 data for the following image analysis.

The reference PSF was used to deconvolve the science image using the Richardson-Lucy deconvolution which converged
within 50 iterations. In addition, we performed reference star differential imaging (RDI) on the IRDIS data. This is
difficult, because the bright central component of the science target is at least marginally extended. As a
consequence, our reference PSF obtained by observing a point-like star is not an ideal representation of the flux
distribution in our science image. Scaling the PSF to the peak counts of the science target and subtracting it reveals
the extended core of the science target, but results in imperfect removal of the core's PSF from its surroundings. To
reach a better contrast further away from the peak emission, we scaled the reference PSF to the same source counts as
the science target in a squared $40\times40$ pixel box (equivalent to $490\times490$\,mas, smaller than the AO
control radius of 800\,mas). This results in a strong oversubtraction of the core brightness, but a better removal of
the core emission from its surroundings. In both cases, the subtraction of the reference PSF leaves strong residuals
due to imperfect removal of the Airy rings, which are more blurred for the science target due to its extended core. The
reduced IRDIS images are shown in Fig.~\ref{fig_irdis_obs}.

For the \textbf{IFS} data, the pipeline was used to create master background and flat field frames, a bad pixel map,
and to determine the wavelength calibration from the observatory provided calibration frames as well as to extract the
spectral image cubes. We then reduced the data for each spectral bin analogous to the IRDIS data. The source structures
in the resulting image are consistent with those in the ZIMPOL and IRDIS images. However, the S/N in these data is
significantly worse than in the IRDIS data. It is dominated by residuals from the data reduction, likely due to
cross-talk, such that combining multiple channels does not result in significant improvements. We thus do not further
consider these data in the present work.

\subsection{Photometric calibration}
\label{sect_phot_cal}
We performed the photometric calibration of our images using the observed PSF reference stars as photometric calibrators.
Aperture photometry of the \textbf{IRDIS} data was performed on each single frame. From the scatter of the counts from those
measurements we estimated an uncertainty of 5\%. We further add in quadrature a conservative uncertainty of 10\% since with
one calibrator observation we are not able to correct for various systematics related to air mass differences, variable
transmission (thin cirrus during the observations), chromaticism, etc. We find magnitudes of AR\,Pup of $6.4\pm0.2$\,mag and
$6.2\pm0.2$\,mag in the H2 and H3 filters, respectively, which is consistent within the uncertainties with the 2MASS
$H$~magnitude of $6.824\pm0.042$\,mag \citep{skr06} and the object's red near-infrared color ($J=7.891\pm0.029$\,mag,
$K_s=5.285\pm0.020$\,mag, not considering variability\footnote{2MASS observations in the $J$, $H$, and $K_s$ band were
obtained simultaneously, so that colors derived from these measurements are not affected by temporal variability between
measurements in single bands.}, \citealt{skr06}). We used our derived magnitude measurements to calibrate the surface
brightness of the extended emission.

Performing photometry on the \textbf{ZIMPOL} data is complicated by the small FoV and the extended halo visible in both
the science and reference observations.  This cannot be distinguished from a halo around the source expected from imperfect
AO correction and thus has to be considered instrumental.  No such halo is visible in the IRDIS images, which is also
consistent with an AO artifact, in which case its intensity would scale with the inverse of the squared wavelength.
We could assume that the fraction of emission in the halo is the same for the observations of science and reference
target and thus calibrate the total magnitude of
our science target. However, to calibrate the surface brightness we need the total source counts. We thus separated the
core and halo emission and approximated the halo with a circular Gaussian. From this, we estimated the total counts in the
halo (inside and outside the FoV) and the total source counts. These could then be used to calibrate the magnitude of
AR\,Pup and the surface brightness of the extended emission. We find that about 45\% and 35\% of the emission are in
the halo in the $V$ and $I$ bands, respectively, and 50\% of the halo emission is outside the FoV in both bands. The
counts from different frames suggest an uncertainty of about 10\%. We conservatively add in quadrature 25\% uncertainty
which is the fraction of total flux that we estimate to be outside the FoV. The resulting brightness of AR\,Pup is
estimated to be $10.1\pm0.3$\,mag and $9.0\pm0.3$\,mag in the $V$ and $I$ bands, respectively. This compares to a $V$
magnitude of 9.6 at the time of the SPHERE observations estimated from our variability monitoring analysis
(Sect.~\ref{sect_observables}).  The $\sim2\sigma$ difference is on the high end of what can be expected by random
measurement errors and the fainter ZIMPOL measurement may suggest that we underestimate the emission in the extended
halo.  This illustrates the difficulties of performing accurate absolute photometry on AO data, in particular with a very
small field of view.

\section{Imaging results}
\label{sect_images}
AR\,Pup's disk is only the second bona-fide post-AGB binary disk that has successfully been imaged in the visible and
near-infrared. Thus the disk geometry and the system's spatial features inferred from our images provide important insights;
not only into the AR\,Pup system, but also the whole class of objects.

\subsection{General morphology}
The reduced images from our observations of AR\,Pup and the reference stars are shown in Figs.~\ref{fig_zimpol_obs}
and~\ref{fig_irdis_obs} for ZIMPOL and IRDIS, respectively. The stacked ZIMPOL images in both filters show a bright
core that is elongated in the North-East to South-West direction. The same elongation can be surmised in the IRDIS
data. Deconvolution of the ZIMPOL images reveals a clear `double bowl' structure separated by a dark band.
This is similar to the appearance of edge-on protoplanetary disks. We thus interpret the dark band as the
optically thick disk mid-plane and the two `bowls' as the two disk surfaces scattering the star light. In the $V$~and $I$~bands
we see no indication of direct stellar light reaching us
through the disk (no central point source is visible).  In $H$~band the lower angular resolution prevents us from drawing a firm
conclusion, but the elongated core of the emission in Fig.~\ref{fig_irdis_obs} suggests that transmitted star light only contributes
a small fraction to the total brightness.  Ancillary optical long baseline interferometric data from the Precision Integrated
Optics Near-Infrared ExpeRiment (PIONIER) at the Very Large Telescope Interferometer (VLTI) show that the contribution of direct
stellar light can be at most 10\% of the total source brightness in the $H$~band (Appendix~\ref{app_pionier}).

The South-Eastern disk surface is brighter than the North-Western one. This suggests that the South-Eastern surface
is oriented toward us, so that the star light scattered on this surface reaches us directly. The light scattered on the
North-Eastern disk surface is partly blocked by the disk following this interpretation. The curvature of the disk mid-plane
also supports this view. The same structures are visible in the IRDIS data after deconvolution or PSF subtraction, albeit
less clear due to the lower angular resolution and stronger residuals from the post-processing. The described features are
the clearest in the deconvolved $I$~band image despite the nominally lower resolution compared to the $V$~band, which we
attribute to a better AO correction at longer wavelengths. This is supported by the images of the reference star, where Airy
rings are marginally visible in the $I$~band, but not in the $V$~band. The dark disk mid-plane has a radius of $\sim$50\,mas,
while bright structures extend up to 300\,mas from the disk center in the deconvolved ZIMPOL images. The size in our IRDIS
images is somewhat smaller, most likely because the higher noise and residuals from PSF subtraction or deconvolution begin
to dominate at smaller separations than in the ZIMPOL images.  The extended halo in the ZIMPOL images has to be attributed to
imperfect AO correction as discussed in Sect.~\ref{sect_phot_cal}.

\begin{figure}
 \centering
 \includegraphics[angle=0,width=\linewidth]{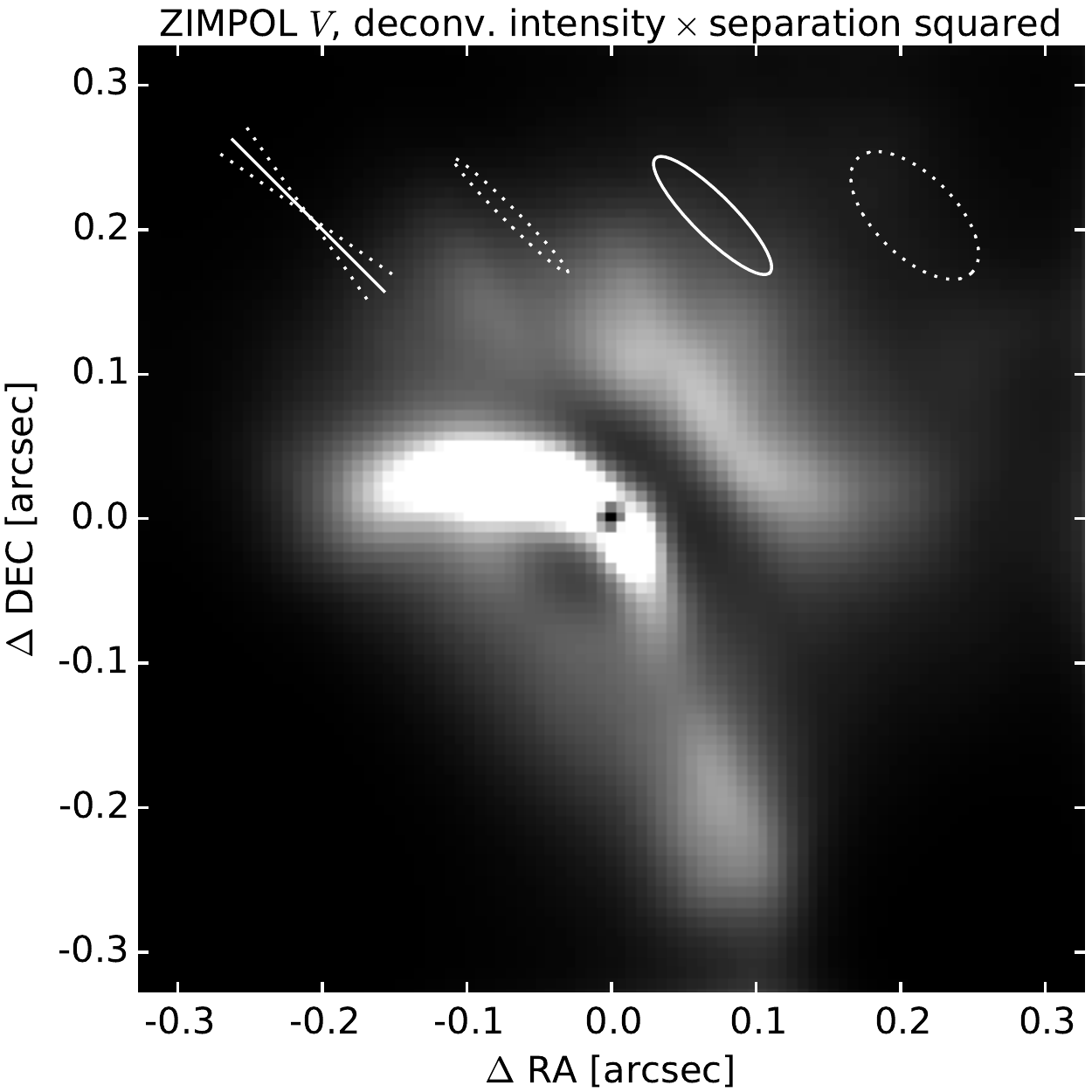}
 \caption{Analysis of the disk's position angle and inclination. The counts in the deconvolved $V$~band image have been
  multiplied by the squared projected separation from the peak location to approximately correct for the geometric
  dilution of star light before being scattered by a dust grain. The solid line and ellipse illustrate the most
  plausible position angle and inclination, while the dashed lines and ellipse illustrate the plausible ranges.}
 \label{fig_inc_pa}
\end{figure}

\begin{figure*}
 \centering
 \includegraphics[angle=0,width=\linewidth]{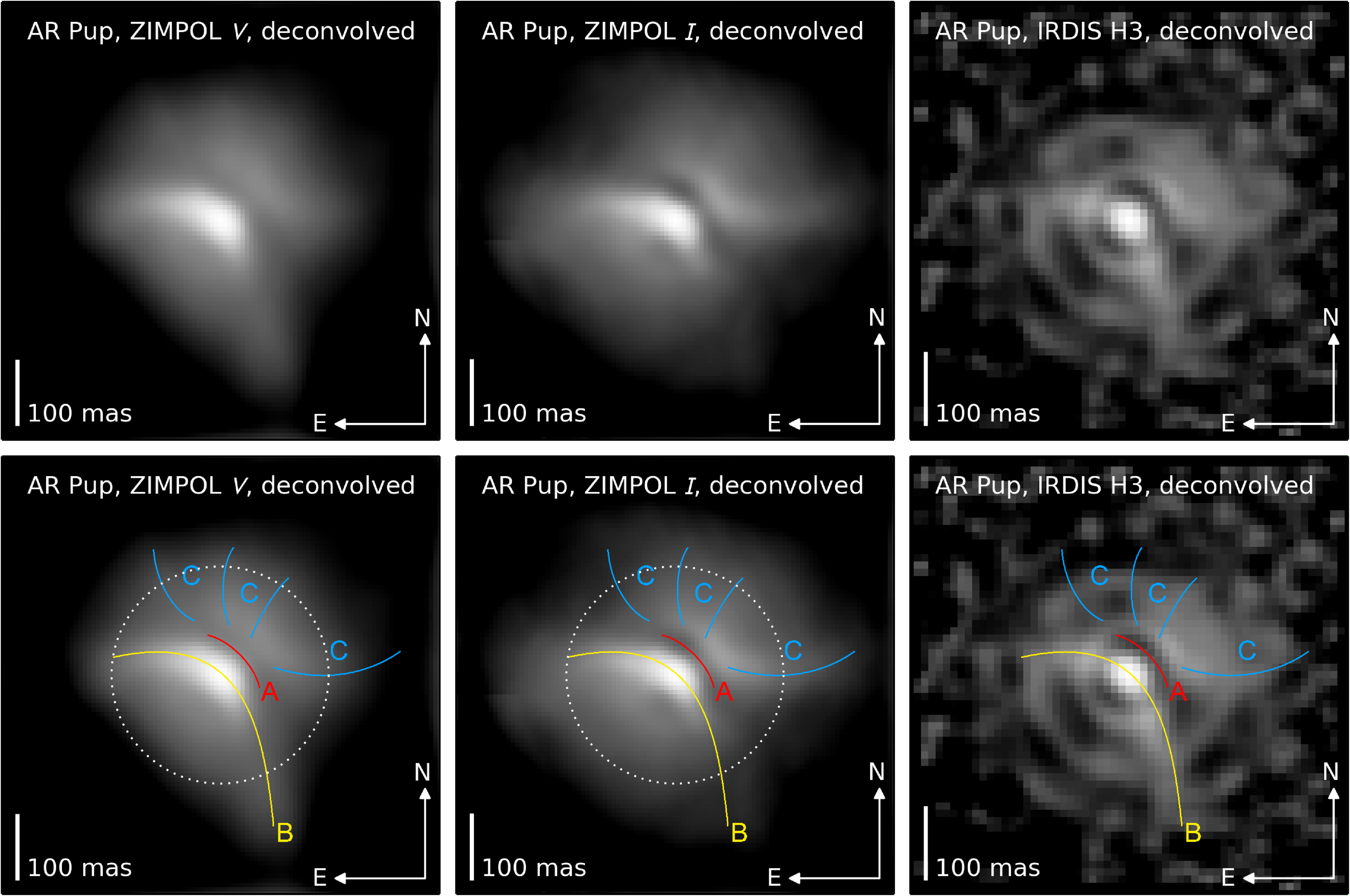}
 \caption{Disk morphology and structures. The deconvolved ZIMPOL $V$ (\emph{left}) band, ZIMPOL $I$ band (\emph{center}), and
  IRDIS (\emph{right}) images are shown in the top row for clarity. The same images are shown in the bottom row with
  overlays highlighting the disk features discussed in Sect.~\ref{sect_substruct}.  The dotted circles marks the separation
  at which the azimuthal profiles in Fig.~\ref{fig_azprof} have been extracted.}
 \label{fig_structures}
\end{figure*}

\subsection{Disk orientation}
We estimate the disk inclination and position angle from the deconvolved
$V$~band image which provides the highest angular resolution. To compensate approximately for the geometric dilution
of the star light before it is scattered by a dust grain at a given distance form the star, we multiply the counts in the
images by the square of the projected separation from the peak location of the emission (the approximate center of the source).
This further highlights the disk mid-plane. The result is shown in Fig.~\ref{fig_inc_pa}. We then qualitatively fit ellipses to
the mid-plane to assess the plausible ranges for the inclination and position angle of the disk. We estimate a disk inclination
of $75^{+10}_{-15}$\,deg from face-on and a position angle of the major axis of $45\pm10$\,deg east of north.  A more formal
analysis requires detailed radiative transfer modeling due to the complex interplay of disk orientation, scale hight, flaring,
optical depth, and scattering phase function of the dust, which is beyond the scope of this paper.

\subsection{Disk substructure}
\label{sect_substruct}
\begin{figure}
 \centering
 \includegraphics[angle=0,width=\linewidth]{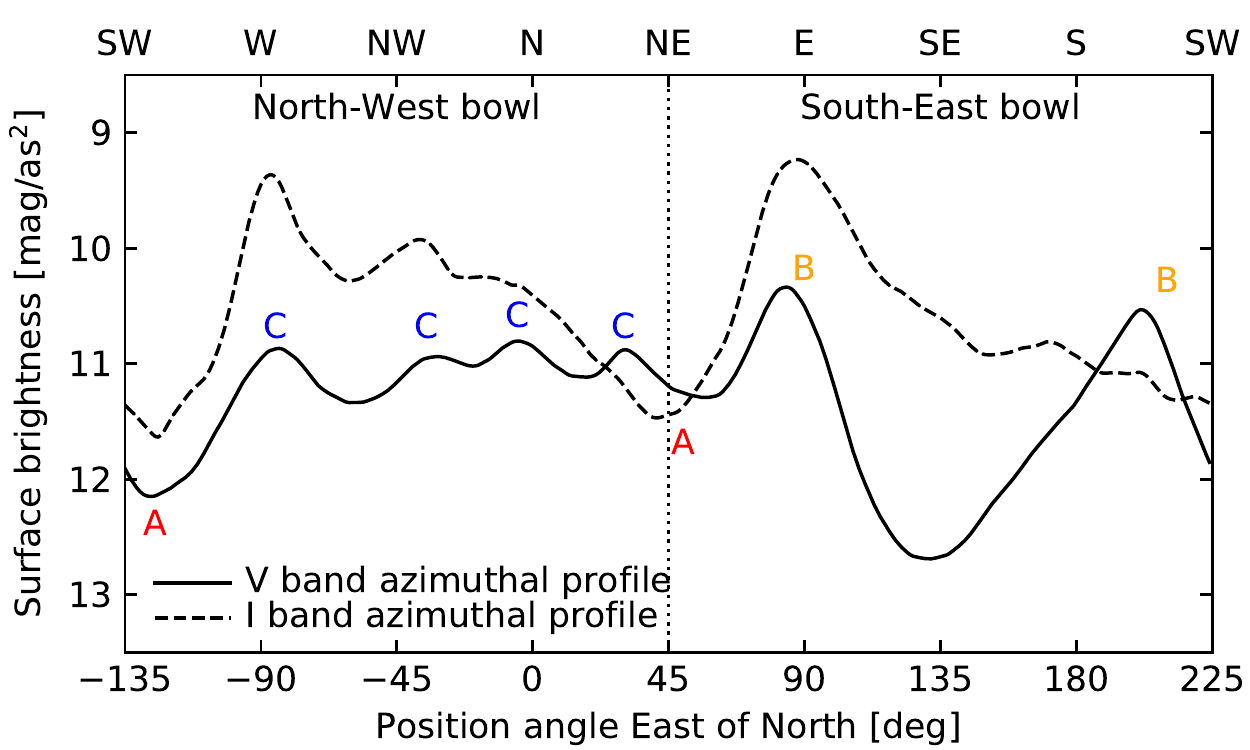}
 \caption{Azimuthal profiles of the images in the two ZIMPOL images at a separation of 175\,mas (25\,pixels) from the location of
 the peak brightness (marked by the dotted circles in Fig.~\ref{fig_structures}).  The letters A, B, and C mark the locations of the
 disk features discussed in Sect.~\ref{sect_substruct} and shown in Fig.~\ref{fig_structures}.  Uncertainties on the profiles are
 dominated by systematics from deconvolution and likely considerable (see Sect.~\ref{sect_substruct} for details).}
 \label{fig_azprof}
\end{figure}

In addition to the global features of the disk image discussed in the previous sections, there are a number of arc-like
substructures visible in both deconvolved ZIMPOL images and in part in the non-deconvolved ZIMPOL and deconvolved
and PSF subtracted IRDIS
images. We highlight the structures visible in the disk in Fig.~\ref{fig_structures}. To further illustrate the discussed
features, we show in Fig.~\ref{fig_azprof} azimuthal profiles of the images in the two ZIMPOL filters at a separation of 175\,mas
(25 pixels) from the location of the peak brightness. The dark mid-plane is marked as feature~A.  The South-Eastern bowl
(feature~B) has a sharp, bright, arc-like contour which could be caused by a combination of geometric and optical effects (viewing 
angle along the surface of the disk oriented toward Earth, forward scattering of the light on large dust grains compared to the
observing wavelength).  The brightest part of the arc shows a strong asymmetry and seems more extended toward the East in both ZIMPOL
images. This could be explained by actual disk asymmetry or by an illumination effect due to the offset of the bright post-AGB star
from the disk center on the binary orbit (see Sect~\ref{sect_rvb} for a more detailed discussion).  Toward the South, the arc
extends into a long, fainter arm in the ZIMPOL $V$~band and IRDIS images. In the ZIMPOL $I$~band image the faint arc seems similarly
extended toward the South.  However, the arc appears fainter in the $I$~band, so that it partly
blends with the rest of the disk emission.  This explains why the Southern peak~B in Fig.~\ref{fig_azprof} is only visible in
the $V$~band profile, not the $I$~band.

The North-Western bowl shows multiple, arc-like substructures in the ZIMPOL images (features~C). These structures also have
tentative counterparts in the deconvolved IRDIS image, although they cannot be well distinguished from deconvolution residuals.
They are weak modulations of the surface brightness and are not obvious in the non-deconvolved images. Their locations
seem anti-correlated with the directions of two PSF features best seen in
the $I$~band reference star image (beams in both directions of the PSF peak at position angles of $\sim$120\,deg and
$\sim$160\,deg East of North). It is possible that these features were not entirely stable during the observations
of science target and reference star. This could have been overcompensated for during the deconvolution process,
resulting in dark artifacts at the location of the PSF features, thus causing apparently bright features at other
locations. However, the PSF features are symmetric around the core of the PSF. The fact that the supposed disk features
are only visible on one side makes this scenario unlikely.

Understanding the origin of these arc-like features in the North-Western side of the disk can give important insights into
the disk architecture, the system's evolution, and binary-disk interaction. \citet{ker15} observed similar features in their
images of the L$_2$\,Pup disk (dubbed plumes in their paper). They suggested the features originate in an outflow of material
from the binary star (wind or jet, as observed, e.g., by \citealt{bol17} in the BD+46\,442 system). An outflow in the form of a
disk wind as has been observed by \citet{buj18} for IRAS\,08544-4431 would be another possible explanation.  If the features are
indeed caused by outflows, they could trace the processes producing the bipolar morphologies observed in more evolved protoplanetary
nebulae and planetary nebulae.

Similar scattered light features have been observed in young protoplanetary disks \citep{cas12, sto16, ben17}. These data
reveal both spirals in the outer disk and structures in the inner disk regions casting shadows onto the outer disk.
\citet{pri18} modeled the binary-disk interaction for the circumbinary disk HD\,142527 and were able to reproduce all
observed features.  These studies can inform similar studies of the features seen in the AR\,Pup and L$_2$\,Pup images.

Monitoring the time evolution of the features can help establishing their nature \citep{deb17}: A slow evolution on the
orbital time scale of material in the outer disk would suggest actual structures in the outer disk. Faster evolution on the
orbital time scale of the disk material close to the star would suggest a shadowing effect. An outflow from the binary is
expected to cause a modulation of the features on the binary orbital time scale.

\section{Discussion}
\label{sect_discuss}

\subsection{Comparison to other disks around evolved stars imaged in scattered light}
\label{sect_comp_disks}

The Red Rectangle is indubitably the case most similar to AR\,Pup of a disk imaged around an evolved star \citep{ost97, coh04}.
The disk around the Red Rectangle's post-AGB RVb binary central star \citep{wae96} could be imaged with the Hubble Space Telescope
due to its larger angular size of $\sim1''$.  It can be explained by a closer distance ($\sim500$\,pc, \citealt{vanle07}), a
larger physical disk size, or a combination of both.
This disk shows a similar over-all shape with a double-bowl structure and a dark mid-plane, but the two bowls show a weaker
brightness asymmetry, suggesting that the disk is seen closer to edge-on.  The primary is of spectral type B9 suggesting that
it is more evolved than AR\,Pup.  The system shows extended red emission which we would be unable to detect in our data of
AR\,Pup due to our small field of view (ZIMPOL) and limited sensitivity to faint, extended emission (IRDIS).  No Hubble images
of AR\,Pup are available.

The disks around L$_2$\,Pup \citep{ker15} and BP\,Psc \citep{zuc08, deboe17} are other potentially similar systems, albeit less clear: 
L$_2$\,Pup has been characterized as a nearby AGB star ($d=64$\,pc, \citealt{vanle07}) with a planetary mass companion
\citep{ker16, hom17} and its disk is more compact with a size of $\sim$25\,AU.  The distance of BP\,Psc has recently been measured
by Gaia to be $\sim$350\,pc \citep{gaia18}, confirming its nature as a first ascent giant star \citep{zuc08}.  The disk has a
diameter of $\sim$35\,AU ($\sim$100\,mas at $\sim$350\,pc), similar to that of L$_2$\,Pup.  Neither L$_2$\,Pup nor BP\,Psc are
known to show long-term variability that could be attributed to binary orbital motion.

If the planetary mass companion of L$_2$\,Pup was responsible for the formation of the disk, it cannot have delivered as much angular
momentum as the stellar mass companion of AR\,Pup.  This would explain the smaller disk around L$_2$\,Pup and the fact that the
star is still characterized as an AGB star, i.e., the interaction with the companion that created the disk did not terminate the
star's AGB phase.  The disk size and evolutionary state of BP\,Psc as a first ascent giant star \citep{zuc08} suggest that for
this star, too, a companion of planetary mass  --  rather than stellar mass  --  may be responsible for the disk formation.

Given their small angular sizes compared to the angular resolution of the instrument with which they were imaged, it is no
surprise that all four disks are seen close to edge-on, at least partly obscuring their host stars.  Imaging disks oriented
closer to face-on requires more sophisticated high contrast techniques to reveal circumstellar disks of similar size, such as
polarimetric differential imaging.  Such observations of post-AGB disks have the potential to provide more
insight into the disk properties as they better allow us to image radial and azimuthal brightness variations of the disk surface
more directly.

\subsection{Connection to other observables of the system}
\label{sect_observables}

We have demonstrated that it is now possible with the newest extreme AO systems to image disks around post-AGB binaries. This
is, however, typically still only possible for systems within a few kpc -- depending on linear disk size. Furthermore, AR\,Pup
is particularly well suited for this due to its edge-on disk masking the bright stellar light, thus reducing the required
image contrast.  Spatially resolved imaging of a few benchmark systems such as AR\,Pup and connecting other observables of these
systems to the properties of the disks is vital to our understanding of systems that cannot be resolved.  Here we briefly discuss
AR\,Pup's SED and photometric variability in the light of our SPHERE images.  A more comprehensive analysis will be enabled by
imaging a sample of post-AGB binary disks.

\subsubsection{Spectral Energy Distribution}

\citet{hil17} found from their study of AR\,Pup's SED that the total infrared luminosity is higher than that derived from the
dereddened optical fluxes.  They suggest that the disk of AR\,Pup must be seen close to edge-on, so that most direct star light
is blocked.  We confirm this conclusion and expand on it with our resolved images of the system, showing that the light seen at
visible wavelengths stems entirely from star light scattered in our direction by the dust on the disk surface.  The derived value
of $E(B-V)$ has thus to be understood as the result of a combination of effects \citep{sci15}:  The light is altered by scattering
events in the disk and affected by circumstellar and interstellar extinction on its \emph{indirect} path from the star toward us.
The measured $E(B-V)$, however, heavily underestimates the circumstellar extinction on the \emph{direct} line of sight from the
star.  Thus, the unreddened stellar model in Fig.~\ref{fig_sed} heavily underestimates the total stellar emission, so that it
cannot be used to determine the stellar luminosity or the star's distance if the luminosity is estimated or derived independently.

At near-infrared and longer wavelengths we start seeing the thermal emission of the disk. The onset of this emission around 
the $H$~band (Appendix~\ref{app_pionier}) indicates that the hottest material has a temperature of $\sim$1200\,K. The bulk
emission peaks around 10\,$\mu$m, indicating a temperature around $\sim$550\,K for the majority of the dust.  Detailed radiative
transfer modeling of our scattered light images and the SED can be used to derive accurate disk parameters such as geometry and dust
properties.

\begin{figure}
 \centering
 \includegraphics[angle=0,width=\linewidth]{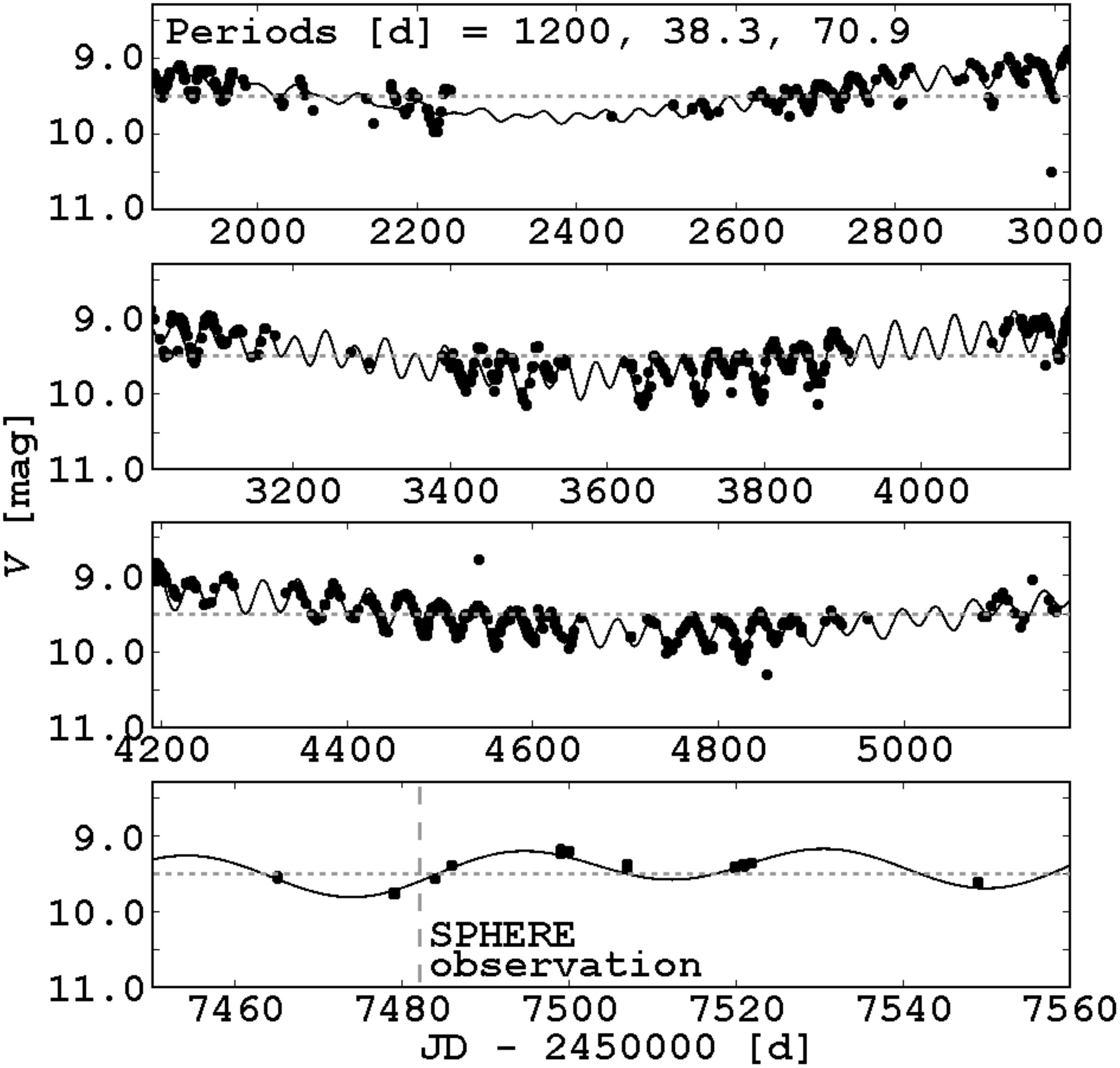}
 \caption{Photometric monitoring of AR\,Pup. The horizontal, dotted line indicates the approximate mean magnitude of $V=9.5$.
   The time of the  SPHERE observations is highlighted by the vertical, dashed line.}
 \label{fig_variability}
\end{figure}

\subsubsection{RVb phenomenon}
\label{sect_rvb}

The fact that no direct stellar light is seen in the visible also has consequences for the interpretation of the system's
observed RVb variability.  To accurately determine the phase of the variability during our SPHERE observations, we complement
photometric time series data
from the All Sky Automated Survey (ASAS, \citealt{poj02}) with own photometric monitoring observations between 2016~March~17
and 2016~June~9 (around the time of our SPHERE observations). Our observations were performed at Mt.~Kent Observatory (near
Toowoomba, Queensland, Australia) using a PlaneWave CDK700 telescope equipped with an Alta U16M Apogee camera. Data reduction
and photometric analysis were performed using the AstroImageJ software package \citep{col17}.  We analyze the ASAS and Mt.~Kent
data following \citet{man17} and recover the 76.66\,days pulsation period and the 1194\,days RVb period found by \citet{kis17}
within the uncertainties. We find that the brightness of AR\,Pup was increasing and close to its average value during our SPHERE
observations from both the pulsation variability ($\sim$5~days past minimum) and the long-term variability ($\sim$300~days past
minimum). This is illustrated in Fig.~\ref{fig_variability}.  Imaging the system at different phases of its variability
will allow us to connect the photometric variation, variations in the disk features, and the orbital phase of the binary.

The origin of the long-term variability of RVb stars has been attributed to either variable extinction or variable scattering of
star light by a circumbinary disk.  For example, \citet{wae91} concluded for HR\,4049 based on the color dependence of the
variability that it likely stems from variable extinction.  For the Red Rectangle, \citet{wae96} concluded that it must stem from
variable scattering because they don't see any color variation and the star is completely obscured by the edge-on disk. 
\citet{kis17} analyzed the amplitude variations of the short period variability of RVb stars over the long period and conclude
that they are consistent with variable obscuration by the disk.

For AR\,Pup, we show that the star is fully obscured by the disk, similar to the Red Rectangle.  The star was close to its mean
brightness in both the long-term and short-term variability cycles, so we cannot simply see the moment of maximum obscuration
over the cycle of variable extinction.  Instead, the variability must be caused by variable scattering on the disk surface,
similar to the Red Rectangle.  We find the disk of AR\,Pup to be slightly inclined away from edge-on and we see structures in the
disk that may be attributed to illumination effects.  We thus propose that variable disk illumination over the binary's orbital
period is an alternative scenario (although the two are not mutually exclusive) to the variable scattering angle as the main
source of the brightness variations proposed by \citet{wae96} for the Red Rectangle.

\begin{figure}
 \centering
 \includegraphics[angle=0,width=\linewidth]{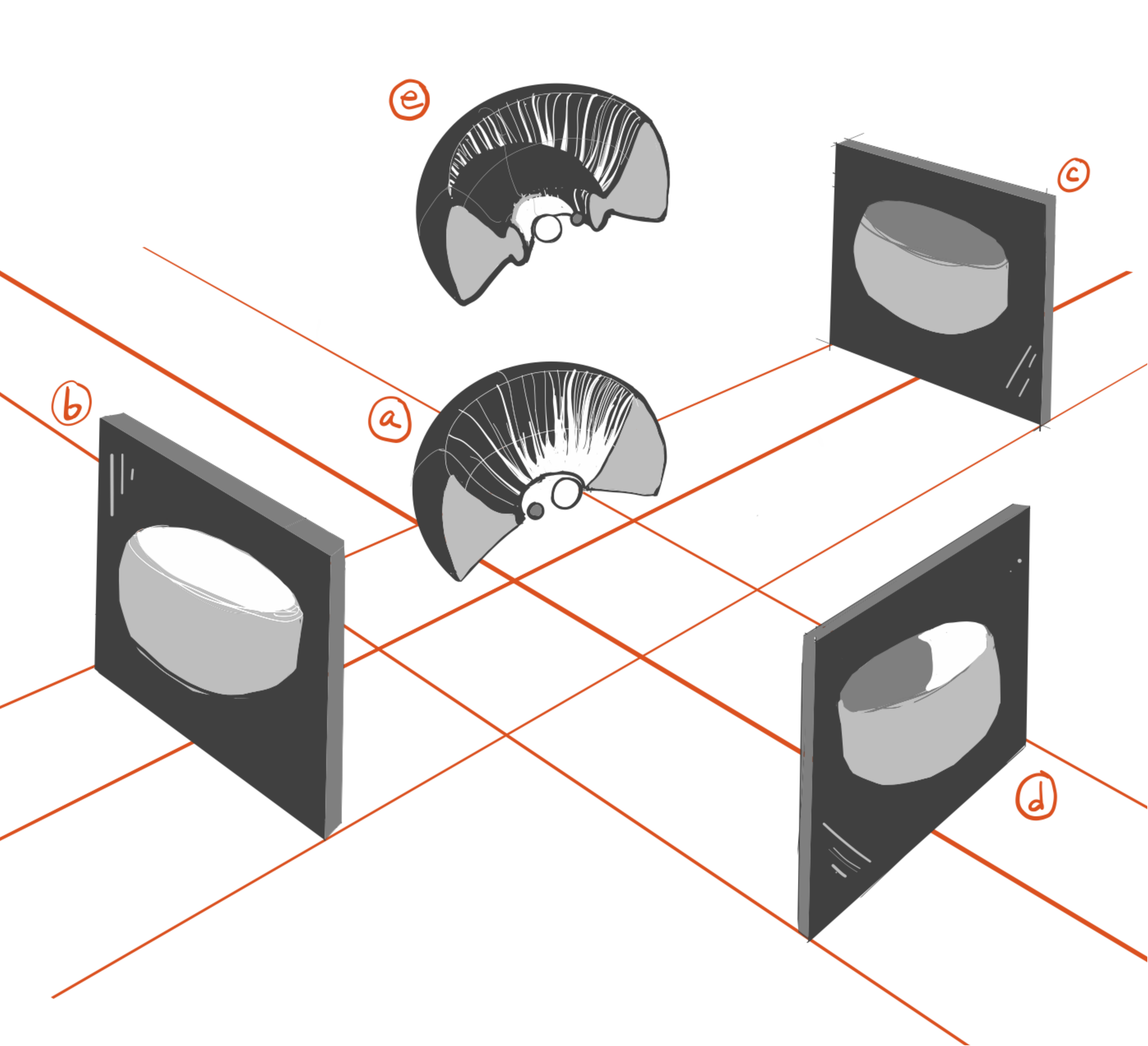}
 \caption{Illustration of the disk illumination effect producing the RVb phenomenon in the AR\,Pup disk.  A disk with a
   vertically thin inner edge is shown (a) and its appearance when viewed from various orientations is illustrated:  (b)
   appearance if the post-AGB star is on the far side of its orbit, (c) appearance if the post-AGB star is on the near side, and
   (d) appearance if the post-AGB star is on the right.  A disk with an inflated inner edge is also shown (e), but the binary
   phase is inverted so that the appearance of the disk in panels (b), (c), and (d) is the same.}
 \label{fig_sketch_rvb}
\end{figure}

We illustrate in Fig.~\ref{fig_sketch_rvb} how variable illumination of the disk by the post-AGB star on its orbit can readily
explain both the brightness variation and the asymmetric brightness of the disk in our images.  We start with a disk that has a
vertically thin inner edge.  The side of the disk closer to the bright post-AGB star on its orbit will be brighter than the other
side.  When the star is on the far side of its orbit, the far, visible side of the disk will be brighter, while the near side of
the disk will be fainter (b).  The system will thus be in a bright phase.  As the post-AGB star moves to the near side on its
orbit, the brightness of the system decreases, as visible side of the disk becomes fainter (c).  During our SPHERE observations,
the star was approximately half way between the near and far sides on its orbit, in which case the disk would show a brightness
asymmetry due to the way it is illuminated (d).  An orbital period of 1194 days implies an orbital semi-major axis of
$\gtrsim1$\,AU. If the inner edge of the disk is plausibly at 5\,AU, then the star would be separated by $\gtrsim4$\,AU from one
side and $\gtrsim6$\,AU from the other side of the disk, causing a difference in illumination by a factor $\gtrsim2.25$.  This
is large enough to explain the long-term
variability of the system.  Alternatively, the inner rim of the disk may be inflated due to the heating from the star (e).  This
way, it would cast a shadow onto the side of the outer disk close to the post-AGB star.  This side would then be fainter having
the opposite effect compared to the scenario with a vertically thin disk edge.  The effect may be enhanced by the variable
heating of the inner disk edge by the star on its orbit with a larger scale height close to the post-AGB star \citep{klu18}. 
Disk structures created by the gravitational interaction with the binary star may play an additional role in the shadowing
scenario.

These two scenarios demonstrate how resolved imaging in combination with detailed knowledge of the binary orbit can help us
understand the structure of the disk and its interaction with the binary star.  Unfortunately, for AR\,Pup a spectroscopic orbit
is not available at the moment.  Obtaining such an orbit is one of the critical next steps in the study of this system.  In
addition, multi-epoch imaging of the disk is important to monitor the variation of the brightness asymmetry.  Comparing the time
scale of this variability with the time scale of the long-term variability is a critical test of the two scenarios described
above.

\subsection{An extreme laboratory for circumstellar disk evolution}
\label{sect_disk_evol}

Several studies have now established the similarities between post-AGB binary disks and protoplanetary disks (\citealt{deruy05,
gie11, hil14, hil15, hil17}, Scicluna et al., in prep.).  Our results on AR\,Pup add to this picture by showing the similar
morphology and scale of this disk from high fidelity images.  In particular the presence of highly processed grains up to
millimeter size found by these works may come as a surprise given the short disk life times ($\sim$$10^4$\,years; \citealt{buj18})
and the fast
evolution of their host stars ($10^2$ to $10^4$\,years, \citealt{ber16}; $10^3$ to $10^4$\,years, \citealt{sah07, ges10}).  Only
recently, \citet{har18} were able to infer the presence of millimeter sized dust grains in the $10^5$\,year young disk around the
protostar TMC1A, while most protoplanetary disks are only observed at an age of $\gtrsim 10^6$\,years (e.g., \citealt{har98}).
Post-AGB binary disks are thus interesting extreme laboratories to study circumstellar disk evolution and to constrain the time
scales of dust grain growth, which is a critical step in the planet formation process.

Due to their short life time, it remains questionable if post-AGB binary disks could actually form planets.  However, large inner
cavities such as the ones often attributed to the presence of planets in transition disks (although there may be other
formation scenarios, \citealt{ale14, tur14}) have been found in both the disks around the post-AGB binary AC\,Her \citep{hil15}
and the Class~I protostar WL\,17 (age of a few $10^5$\,years) \citep{she17}. In particular, the presence of a few large
bodies such as left-over planetesimals from the main sequence might act as a catalyst in post-AGB binary disks \citep{bir12,
gar13, bit15, bir16}.  The dust masses estimated by \citet{deruy05} seem consistent with those of protoplanetry disks
\citep{and13, ans16, ans17, bar16}.

Several other studies have suggested a variety of avenues for second generation planet formation beyond a star's main sequence life
time, suggesting this might not be an uncommon process.  The claimed\footnote{A number of these candidates have been shown to be
dynamically unstable (e.g., \citealt{hor11, wit13, pul18}), which potentially casts doubt on the existence of those claimed planets.}
discoveries of planetary systems orbiting post-common envelope binary stars (e.g., \citealt{lee09, alm13, pul18}) are best explained
by such a scenario \citep{mus13}.  The planets detected orbiting pulsars are another such example \citep{wol92, wan06}.  Recently,
\citet{vanlie18} proposed a scenario how even Earth mass, rocky planets could form in the habitable zones around white dwarfs.

\section{Conclusions}
\label{sect_conc}

We have presented the first resolved images of the circumbinary disk around the post-AGB star AR\,Pup. Using VLT/SPHERE
observations in the visible and image deconvolution with an observed reference star, as well as reference star differential
imaging, we were able to reveal circumstellar emission in the $V$, $I$, and $H$~bands. The disk is optically thick in all our
images and seen close to edge-on. The dark disk mid-plane has a radius of about 50\,mas and extended emission is detected out
to a separation of about 300\,mas from the source center. We estimate an inclination of $75^{+10}_{-15}$\,deg from face-on
and a position angle of the disk major axis of $45\pm10$\,deg East of North.

Several arc-like features are identified in the visible images. We discussed various scenarios to explain these features,
including a disk wind, outflows or jets from the central binary due to stellar wind or accretion, and shadows cast onto the
outer disk by disk structures closer in due to binary-disk interactions. Resolved images at various phases of the binary
orbit can potentially allow us to distinguish between these scenarios.

There is no indication of direct stellar light in our visible images and likely neither in the $H$~band image.  We thus conclude
that the long-term RVb photometric variability of the system must be caused by variable scattering, not extinction, of star light
over the binary orbit.  We propose a scenario of variable disk illumination in addition to previously discussed variations of the
scattering angle to explain this variability.

Finally, we highlight the value of post-AGB binary disks as extreme laboratories to study circumstellar disk evolution. The
high degree of dust processing observed in these disks despite their short life times of $<10^5$\,years strongly constrains
the time scales at which dust processing in disks can happen. If a connection to protoplanetary disks can be made, this
constrains the swiftness of the early stages of the planet formation process. Along these lines, we argue that second
generation planet formation beyond a star's main sequence life time might not be uncommon.

\acknowledgments

SE acknowledges support through the ESO fellowship program and thanks C.~Melo for his support and encouragement to pursue this
research.
GHMB acknowledges funding from the European Research Council (ERC) under the European Union's Horizon 2020 research and innovation
programme (grant agreement No. 757957).
This research has been supported by the Ministry of Science and Technology of Taiwan under grants MOST104-2628-M-001-004-MY3 and
MOST107-2119-M-001-031-MY3, and Academia Sinica under grant AS-IA-106-M03.
EV is supported by Spanish grant AYA 2014-55840-P.
We have made use of the SIMBAD database, operated at CDS, Strasbourg, France, and NASA's Astrophysics Data System.
\textit{Herschel} is an ESA space observatory with science instruments provided by European-led Principal Investigator consortia and
with important participation from NASA.
This work benefitted from the FEARLESS collaboration (FatE and AfteRLife of Evolved Solar Systems, PI: S.~Ertel).

%

\vspace{5mm}
\facilities{VLT:Melipal (SPHERE).}


\software{Python, Astropy \citep{ast13}, NumPy \citep{oli06}, SciPy \citep{jon01}, and Matplotlib \citep{hun07} libraries.}

\bibliographystyle{aasjournal}
\bibliography{bibtex}

\appendix

\section{SED data}
\label{app_sed}

In Table~\ref{tab_sed} we list the photometry and corresponding references used to build the SED in Fig.~\ref{fig_sed}.

\begin{table}[h]
\caption{Photometry of AR\,Pup.}
\label{tab_sed}
\centering
\begin{tabular*}{0.8\linewidth}{c@{\extracolsep{\fill}}cccl}
\toprule
 Wavelength & Flux    & Uncertainty  & Instrument/     & Reference            \\[-5pt]
 [$\mu$m]   & [Jy]    & [Jy]         & Filter          &                      \\
\midrule
0.34    &    0.060    &    0.012     &    GENEVA       &    \citet{mer97}     \\[-5pt]
0.35    &    0.0495   &    0.0049    &    STROMGREN    &    \citet{hau98}     \\[-5pt]
0.36    &    0.0502   &    0.0050    &    JOHNSON      &    \citet{mer97}     \\[-5pt]
0.40    &    0.252    &    0.049     &    GENEVA       &    \citet{mer97}     \\[-5pt]
0.41    &    0.235    &    0.024     &    STROMGREN    &    \citet{hau98}     \\[-5pt]
0.42    &    0.326    &    0.053     &    GENEVA       &    \citet{mer97}     \\[-5pt]
0.44    &    0.313    &    0.094     &    USNOB1       &    \citet{mon03}     \\[-5pt]
0.44    &    0.317    &    0.032     &    JOHNSON      &    \citet{and12}     \\[-5pt]
0.44    &    0.222    &    0.022     &    JOHNSON      &    \citet{mer97}     \\[-5pt]
0.44    &    0.323    &    0.010     &    JOHNSON      &    \citet{kha07}     \\[-5pt]
0.45    &    0.432    &    0.084     &    GENEVA       &    \citet{mer97}     \\[-5pt]
0.47    &    0.396    &    0.064     &    STROMGREN    &    \citet{hau98}     \\[-5pt]
0.54    &    0.75     &    0.15      &    GENEVA       &    \citet{mer97}     \\[-5pt]
0.55    &    0.781    &    0.097     &    GENEVA       &    \citet{mer97}     \\[-5pt]
0.55    &    0.691    &    0.069     &    STROMGREN    &    \citet{mer97}     \\[-5pt]
0.55    &    0.573    &    0.093     &    STROMGREN    &    \citet{hau98}     \\[-5pt]
0.55    &    0.444    &    0.044     &    JOHNSON      &    \citet{mer97}     \\[-5pt]
0.55    &    0.562    &    0.014     &    JOHNSON      &    \citet{kha07}     \\[-5pt]
0.55    &    0.559    &    0.056     &    JOHNSON      &    \citet{and12}     \\[-5pt]
0.58    &    0.92     &    0.18      &    GENEVA       &    \citet{mer97}     \\[-5pt]
0.65    &    0.499    &    0.050     &    COUSINS      &    \citet{and12}     \\[-5pt]
0.69    &    0.69     &    0.21      &    USNOB1       &    \citet{mon03}     \\[-5pt]
0.79    &    0.732    &    0.073     &    COUSINS      &    \citet{and12}     \\[-5pt]
1.24    &    1.090    &    0.029     &    2MASS        &    \citet{skr06}     \\[-5pt]
1.65    &    1.936    &    0.075     &    2MASS        &    \citet{skr06}     \\[-5pt]
2.16    &    5.140    &    0.095     &    2MASS        &    \citet{skr06}     \\[-5pt]
4.29    &    72.7     &    6.8       &    MSX          &    \citet{ega03}     \\[-5pt]
4.35    &    81.0     &    7.5       &    MSX          &    \citet{ega03}     \\[-5pt]
8.48    &    114.3    &    4.7       &    MSX          &    \citet{ega03}     \\[-5pt]
11.0    &    129.9    &    5.3       &    IRAS         &    \citet{mos90}     \\[-5pt]
11.0    &    131.0    &    7.9       &    IRAS         &    \citet{hel88}     \\[-5pt]
12.2    &    118.0    &    5.9       &    MSX          &    \citet{ega03}     \\[-5pt]
14.7    &    100.2    &    6.1       &    MSX          &    \citet{ega03}     \\[-5pt]
21.5    &    91.5     &    5.5       &    MSX          &    \citet{ega03}     \\[-5pt]
23.1    &    90.2     &    4.3       &    IRAS         &    \citet{mos90}     \\[-5pt]
23.1    &    94.3     &    4.7       &    IRAS         &    \citet{hel88}     \\[-5pt]
58.2    &    24.3     &    1.5       &    IRAS         &    \citet{mos90}     \\[-5pt]
58.2    &    26.1     &    2.6       &    IRAS         &    \citet{hel88}     \\[-5pt]
65.4    &    20.13    &    0.92      &    AKARI        &    \citet{mur07}     \\[-5pt]
85.1    &    14.74    &    0.28      &    AKARI        &    \citet{mur07}     \\[-5pt]
99.5    &    12.0     &    1.1       &    IRAS         &    \citet{hel88}     \\[-5pt]
99.5    &    12.4     &    2.2       &    IRAS         &    \citet{mos90}     \\[-5pt]
146     &    5.012    &    0.85      &    AKARI        &    \citet{mur07}     \\[-5pt]
162     &    4.7      &    1.1       &    AKARI        &    \citet{mur07}     \\[-5pt]
250     &    1.610    &    0.083     &    SPIRE        &    \citep{schu17}    \\[-5pt]
350     &    0.861    &    0.099     &    SPIRE        &    \citep{schu17}    \\[-5pt]
500     &    0.429    &    0.073     &    SPIRE        &    \citep{schu17}    \\
\bottomrule
\end{tabular*}
\end{table}

\section{PIONIER observations}
\label{app_pionier}

\begin{figure}
 \centering
 \includegraphics[angle=0,width=0.5\linewidth]{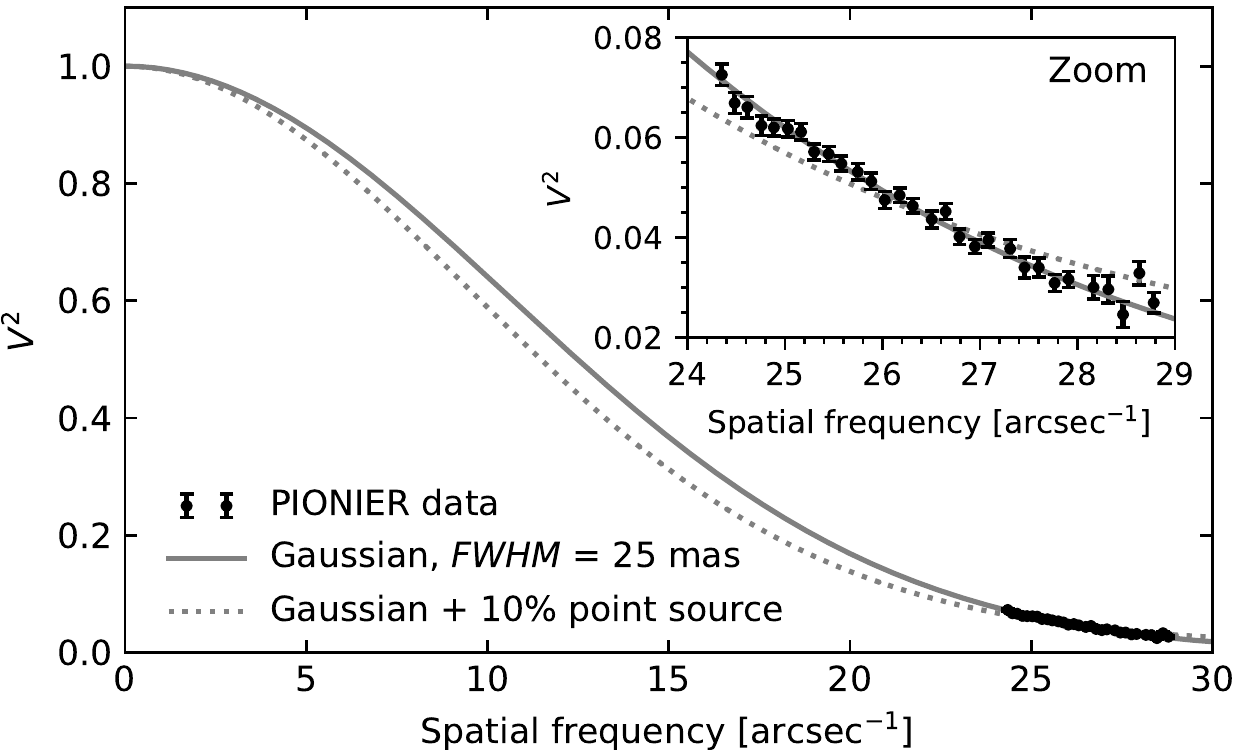}
 \caption{PIONIER squared-visibility measurements of AR\,Pup in $H$~band at a projected changing from 8.9\,m to 9.1\,m over the ten
  minute sequence of exposures.  Overplotted are
  a Gaussian source brightness distribution with $\textsl{FWHM} = 25$\,mas and a model consisting of a Gaussian
  ($\textsl{FWHM} = 29$\,mas) plus a point source contributing 10\% to the total source flux.}
 \label{fig_pionier}
\end{figure}

PIONIER \citep{lebou11} observations of AR\,Pup were carried out on 2015~Feb~24 (Project 094.D-0865, PI: Hillen) using the compact
array configuration of the Auxiliary Telescopes of the VLTI (interferometric baselines between 11.3\,m and 34.3\,m).  Fringes could
only be detected at the shortest baseline (projected baseline changing from 8.9\,m to 9.1\,m over the ten minute sequence of
exposures).  Five
sequences of 100 scans each were recorded in six spectral channels across the $H$~band, resulting in $5\times6 = 30$
squared-visibility ($V^2$) measurements.  HD\,67556 was observed
after AR\,Pup as a calibration star in the same instrument setup.  Data were reduced using version 3.30 of the PIONIER pipeline.
The squared-visibility measurements are shown in Fig.~\ref{fig_pionier}.

The bright central star of AR\,Pup is expected to be unresolved at short baselines (diameter <1\,mas) and the contribution of the
faint companion
is negligible, so that the stellar $V^2$ is expected to be one.  The fact that fringes could only be detected at the shortest
baseline and the low visibility suggest that the source is extended.  We show in Fig.~\ref{fig_pionier} that a Gaussian source
brightness distribution with a full width at half maximum (\textsl{FWHM}) of 25\,mas reproduces the data well.

In order to put an upper limit on the amount of direct star light contributing to the total source emission in $H$~band, we add
to the Gaussian brightness distribution an unresolved point source (the star) following, e.g., \citet{dif07}.  No satisfactory
fit with a substantial amount of direct star light could be achieved.  We show in Fig.~\ref{fig_pionier} a model with a 10\% point
source and an \textsl{FWHM} of the Gaussian component of 29\,mas, which produces a mediocre fit to the data.  We thus put a
conservative upper limit of 10\% on the contribution of direct star light to the total source emission.  The fact that no fringes
were detected at longer baselines also puts an upper limit of $\sim10$\% on the contribution of any point source to the total
emission assuming the disk is fully resolved at these baselines and given that a squared visibility of $1$\% can be detected by
PIONIER.

Our measurements indicate that the contribution of direct star light to the total brightness of AR\,Pup in the $H$~band must be
small.  The strong brightness increase of AR\,Pup from the $J$~band to the $H$~band (Fig.~\ref{fig_sed}) must thus be dominated
by the onset of hot thermal disk emission rather than the disk becoming optically thin and transmitting direct star light.

\end{document}